\documentclass[reprint,eqsecnum,floats,aps,amsmath,amssymb,nofootinbib,prd,onecolumn, showpacs]{revtex4-1}

\usepackage{graphicx}
\usepackage{amsmath,amssymb}
\usepackage{hyperref}
\usepackage{graphicx}
\usepackage{subfigure}
\usepackage{arydshln}
\usepackage{inputenc}
\usepackage{graphicx}
\usepackage{amsmath,amssymb}
\usepackage{hyperref}

\begin{document}
	
\title{Primordial perturbations in kinetically dominated regimes of general relativity and hybrid quantum cosmology}

\author{Beatriz Elizaga Navascu\'es}
\email{w.iac20060@kurenai.waseda.jp}
\affiliation{JSPS International Research Fellow, Department of Physics, Waseda University, 3-4-1 Okubo, Shinjuku-ku, 169-8555 Tokyo, Japan}
\author{Rafael Jim\'enez-Llamas}
\affiliation{Instituto de Estructura de la Materia, IEM-CSIC, Serrano 121, 28006 Madrid, Spain}
\email{rafael.jimenez@iem.cfmac.csic.es}
\author{Guillermo  A. Mena Marug\'an}
\email{mena@iem.cfmac.csic.es}
\affiliation{Instituto de Estructura de la Materia, IEM-CSIC, Serrano 121, 28006 Madrid, Spain}

\begin{abstract}
Scalar fields with an energy density dominated by its kinetic part may have played a relevant role in the very early stages of the Universe. Compared to the standard inflationary paradigm, they may lead to modifications in observable quantities, e.g. the anisotropies found in the cosmic microwave background. Kinetically dominated regimes arise in classical fast-roll scenarios as well as in quantum bouncing cosmologies. For instance, kinetic dominance is typical in interesting preinflationary phases of Loop Quantum Cosmology. In this work, we analyze the leading-order effects that the presence of a scalar field potential causes on the primordial cosmological perturbations  in these kinetically dominated epochs. These effects can be grouped in two sets, namely, those that affect the effective mass of the perturbations and those that affect the choice of their vacuum state. The effective mass is modified directly by terms that include the potential, but also indirectly by the change in the background dynamics and the relation between the parameterization of these dynamics and the conformal time, usually employed to describe the evolution of the perturbations. On the other hand, away from de Sitter inflation, the Bunch-Davies state is no longer the most natural vacuum at all scales. Recent proposals suggest to modify it by carrying out certain Hamiltonian diagonalization with a suitable asymptotic behavior at large wavenumber scales. Both this diagonalization condition and the imposed asymptotic behavior depend on the effective mass of the perturbations, and therefore the selected vacuum state varies in the presence of the scalar field potential.  
\end{abstract}


\maketitle

\section{Introduction}\label{sec:Intro}

The level of precision that we have reached in cosmological observations has opened a new era in which we can falsify the predictions of the standard cosmological model \cite{Cosmology}, which includes a sufficiently large epoch of slow-roll inflation and a Gaussian distribution of perturbative inhomogeneities that can be explained by assigning to them a vacuum identified with the Bunch-Davies state \cite{BunchDavies}. This standard model has proven extremely successful. Nonetheless, some observations have raised the possibility that there may exist tensions with the theory \cite{Crisis,AshtPRLLast,Ashtekarlast,Ivanlast}. For instance, the observations of the Cosmic Microwave Background (CMB) reported by WMAP and the Planck mission \cite{WMAP,Planck,Planck-inf} indicate a lack of power at large angular scales (for multipoles with number $l$ around and below 30). Another example is the lensing amplitude associated with the gravitational lensing experienced by the CMB radiation in their propagation from the surface of last scattering \cite{Lensing}. Although, individually, the statistical significance of these observational anomalies is small, their combination points toward the exceptionality of the observed Universe unless we alleviate the tension with some new physics, beyond the otherwise successful standard inflationary paradigm.

Considerable attention has recently been paid to possible solutions to this tension, in particular to the anomalous lack of power in the CMB, within the framework of classical General Relativity for spatially flat cosmological models \cite{CPKL,CCL,WNg,DVS1,DVS2,BDVS,JCGSS,SGC,NFS,DKPS,KKO,PL,HHHL}. A solution to this problem could come from the realization that pre-inflationary cosmological stages can leave a trace in the observable Universe (see e.g. \cite{Turok}), for instance if the slow-roll period is not too large, although sufficiently long to avoid conflicts with the observational constraints on the number of e-folds \cite{RS,Ramirez,SA}. The simplest situation with this behavior is found in models with a single scalar field in which the short slow-roll period is preceded by an epoch in which the energy density of the scalar field is dominated by its kinetic part, experiencing what has been called a fast-roll inflation \cite{CPKL,DVS1,DVS2,HHHL}. The onset of inflation, shortly before slow-roll conditions hold in this class of models, typically introduces an infrared scale from which one may observe traces of fast-roll inflation on the primordial power spectrum \cite{Bealast}.

Another possibility to alleviate the tensions comes from quantum cosmology, i.e., the description of the cosmological evolution according to quantum principles \cite{QCHalliwell,LQC1}. In quantum cosmological models, the pre-inflationary history of the Universe can be modified by quantum gravity phenomena as one approaches the Big Bang, in some circumstances providing even a bouncing mechanism that avoids this singularity. This is the case, for instance, when one studies certain physical states for the background in Loop Quantum Cosmology (LQC) \cite{APS1,APS2,MMO}.  LQC is a quantum description of cosmological spacetimes \cite{LQC1,LQC2,LQC3,Hybridreview} constructed by applying the techniques of non-perturbative Loop Quantum Gravity \cite{LQG1,LQG2}. In LQC, there exist certain families of Gaussian states that are peaked around trajectories on phase space that never reach the cosmological singularity \cite{LQC1,Taveras,Louko}. When the energy density of the scalar field is large, these trajectories deviate from those of Einsteinian cosmology. In this departure from General Relativity, they experience large quantum geometry effects. These quantum pre-inflationary phenomena may affect the CMB power spectrum \cite{AshtPRLLast,Ashtekarlast,Ivanlast,AAN,AAN1,AAN2,AG,AAG,ABS,AKS2,NO,hybten,hybpred1,hybpred2,hybothers} if the energy density of the scalar field is kinetically dominated in those epochs \cite{IvanPhase}. 

Through the consideration of these families of states, it has been argued that the loop quantization favors a sufficient amount of slow-roll inflation \cite{sloan,barr,barrsu}. This prediction depends on assumptions about the probability measure on the space of inital data. Actually, with some proposals about this measure, it is claimed that the most probable scenario is that the scalar field is in a kinetically dominated regime at the bounce \cite{barr}. For the purposes of our investigations, the viewpoint that we adopt here is a bit more pragmatic, in the sense that we are interested in peak trajectories of phenomenological interest. Namely, they must be compatible with the CMB observations (in what respects the amplitude and the spectral index of the scalar power spectrum) and the contraints on the number of e-folds during inflation, while displaying quantum geometry effects in the window of wavelength scales that can be observed today. For this class of trajectories, indeed, the energy of the scalar field at the bounce happens to be highly dominated by its kinetic contribution \cite{IvanPhase}.

A similar situation, with kinetically dominated regimes that can leave traces in the CMB power spectrum, could be found in other quantum bouncing cosmological models, not necessarily within the context of LQC. A general framework, that allows for more general quantizations of the cosmological background than a loop quantization, is provided by the so-called hybrid quantum cosmology formalism. The hybrid quantization strategy combines such a quantum description of the background with a Fock description of its gauge invariant perturbations. The background quantization should include the construction of a space of states\footnote{In general, these states are non-physical inasmuch as they still have to satisfy quantum constraints. One usually refers to this space of states as the kinematic (Hilbert) space.} and an inner product over the background geometry. At least within this hybrid quantization, it is possible to show that the scalar perturbations, eventually responsible for the temperature anisotropies of the CMB and that are usually described in terms of the Mukhanov-Sasaki (MS) field \cite{Mukhanov,Sasaki,Sasaki2} (which is perturbatively gauge invariant), satisfy similar evolution equations in conformal time as in General Relativity, except for the modification of the effective-mass term that appears in those equations \cite{MSCastello}. In hybrid quantum cosmology, this mass is given by the ratio of the expectation values of two geometric operators, on the state that determines the quantum background. These expectation values are defined with the aforementioned inner product over the geometry. The quantum modifications of the effective mass affect the propagation of the primordial perturbations, making it possible that there appear changes in the primordial power spectrum if inflation is short lived. In particular, the quantization of the background cosmology typically introduces a scale of Planck order in the system, from which imprints of quantum cosmology phenomena may appear on this spectrum \cite{Bealast,IvanPhase}.

It is worth commenting that MS equations for the scalar perturbations with an effective mass modified by quantum geometry effects appear as well  in other formalisms of quantum cosmology, like another approach to LQC called the dressed-metric formalism \cite{AAN1}.\footnote{Certainly, there exist quantum cosmology formalisms in which the quantum corrections to the MS equations cannot be reduced to a modification of the effective mass (see, e.g., the so-called deformed constraint algebra approach \cite{Bojo0,Bojo1,CLB,Bojo2,BM,ebounce}), which escape from the scheme that we consider in this work.} In this case, the background also presents an epoch of kinetic dominance, and the effective mass of the perturbations is given again in terms of a quantum expectation value, but different from the ratio of expectation values of the hybrid approach.

A preliminary study of the dependence of the modified MS dynamics on the scalar field potential, when it is small, was carried out in Ref. \cite{qcCastello}. Nonetheless, apart from being restricted to the hybrid LQC case, this study is incomplete at least in two aspects. First, any of the considered expectation values is subject to the quantum dynamics of the cosmological background, which is usually described in terms of an internal degree of freedom. On the other hand, the MS equations for the perturbations are parameterized in conformal time. The relation between this time and the internal variable used in the background evolution is also affected by the presence of a potential with respect to the situation with a free scalar field. This important point was not considered in Ref. \cite{qcCastello}. 

Moreover, even if we have specific MS equations at our disposal to dictate the propagation of the scalar perturbations, their evolution cannot be integrated unless we fix some initial conditions for them. These initial conditions are usually interpreted as the choice of a vacuum state for the perturbations. In a standard slow-roll inflation that lasts long enough, so that the window of observable scales in the CMB exited the cosmological horizon only well inside the slow-roll regime, the relevant part of the evolution can be approximately described as a de Sitter phase (with small corrections), and in this sense it is natural to adopt initial conditions that correspond to the choice of a Bunch-Davies state for the perturbations \cite{BunchDavies}. Nonetheless, in scenarios of fast-roll inflation, for instance in General Relativity or following quantum bounces, or more generically away from genuine slow-roll schemes, the adoption of a Bunch-Davies vacuum is no longer justified for scales that can be affected by the new physics, compared with the standard model. The vacuum state should be optimally adapted to the new background dynamics, which is not close to de Sitter anymore.

Several proposals have been put forward to select a state with those properties \cite{AG1,AG2,NO,Mercelast}. Among them, it has been recently suggested that the choice of a set of positive frequency solutions, that determines the vacuum state, should come from a process of diagonalization of the contribution of the perturbations to the Hamiltonian of the cosmological system, which involves an extension from the asymptotic sector of modes with infinitely large wavenumber \cite{BeaDiagonal,BeaDiagonal2}. This diagonalization provides quantum excitations of positive frequency that do not interact (at least asymptotically) when the background dependence of the system is conveniently taken into account, and in this sense it is natural to consider the associated solutions as a natural set to define a vacuum. In addition, it has been shown that this vacuum reproduces the Bunch-Davies one in de Sitter \cite{BeaDiagonal} and that, in phenomenologically interesting situations, one can relate it with other proposals to choose a state that minimizes the oscillations in the primordial power spectrum \cite{NO,BeaDiagonal2} (state that is often called the NO-vacuum). The corresponding choice of positive frequencies can be given in terms of the imaginary part of a (mode-dependent) solution to a Riccati equation that includes the effective mass of the perturbations \cite{BeaDiagonal}. The desired solution to this equation must satisfy a specific asymptotic expansion for large wavenumbers, expansion that also depends on the effective mass. Therefore, modifications of this effective mass of the perturbations with respect to the free scalar field case caused by a potential affect the choice of the vacuum, both in those pre-inflationary models in which the background is described classically within General Relativity and in quantum cosmology models. This influence of the potential on the choice of vacuum state, and therefore on the resulting power spectrum, was not studied in Ref. \cite{qcCastello}. 

The aim of this work is to discuss the influence of the potential on the perturbations in regimes that are kinetically dominated, analyzing the leading-order corrections around the case of a free scalar field. Thus, the goal is to reduce all the complications that a scalar field potential introduces on the background dynamics to the free case plus manageable corrections. In this way, one can simplify the complicated calculations required to solve the cosmological models with a generic potential, for which there is no analytic solution available in general, and which would typically need the implementation of numerical computation techniques. Furthermore, the evaluation of the effective mass would have to be performed numerically for each of the possible values that the scalar field may take (since the effective mass depends explicitly on this field, both classically and quantum mechanically), with the subsequent aggravation of the computational problems. In addition, numerical simulations would also be needed to invert the relation between the scalar field and the conformal time\footnote{In the kinetically dominated regimes that we are considering, the potential remains so small that its presence does not break the monotonicity of the evolution of the scalar field, even when this potential is treated exactly.} in order to determine the effective mass in terms of this time, and these simulations would have to be performed independently for each potential that one wants to consider. All these difficulties would complicate a general investigation of the aforementioned pre-inflationary regimes to the extreme. On the contrary, such a study should be feasible with the approximations that we develop in this work if one can just handle the case of a free scalar field.

The rest of this article is organized as follows. In the next section, we summarize very briefly the hybrid approach to quantum cosmology, the modified MS equations for the scalar perturbations, which can be applied to the case of General Relativity in the classical limit,  and the basic results of the asymptotic diagonalization criterion for the choice of a vacuum state. Sec. \ref{modMass} deals with the corrections to the effective mass that appears in the (modified) MS equations owing to the consideration of a scalar field potential. In a first subsection we calculate the leading-order correction to the effective mass in terms of the scalar field, interpreted as the internal degree of freedom with respect to which we describe the background evolution. The analysis of that subsection differs from Ref. \cite{qcCastello} in several convenient choices of conventions and a neater extraction of the non-free part of the evolution of the scalar field. In a second subsection we discuss the relation between the conformal time and the scalar field in our leading-order approximation in the potential. Finally, in a third subsection we combine these results to derive the total leading-order correction to the effective mass of the perturbations, and comment on the particularization of the result to the cases of a classical Einstenian evolution, hybrid LQC, and a hybrid quantum cosmology model based on geometrodynamics \cite{QCHalliwell}. Sec. \ref{modFreq} investigates the leading-order effects of the scalar field potential on the choice of vacuum state if one adopts the asymptotic diagonalization criterion. Finally, Sec. \ref{conclusion} contains the conclusions. We take units such that the Planck reduced constant, $\hbar$, and the speed of light, $c$, are equal to the unit.

\section{Modified perturbation equations and choice of a vacuum}\label{sec:HLQC}

In order to motivate the kind of modification to the effective mass of the scalar perturbations that can be due to quantum geometry effects, let us consider the hybrid approach to quantum cosmology \cite{MSCastello}. We consider a cosmological spacetime of the Friedmman-Lema\^{\i}tre-Robertson-Walker (FLRW) type with a homogeneous scalar field as matter content, subject to a potential. We assume compact spatial sections with the compact topology of the three-torus,\footnote{The non-compact case is obtained by appropriately taking the limit in which a physical length scale of reference is sent to infinity; see Ref. \cite{Beacontinuous}.}  with orthogonal coordinates chosen with period equal to $2\pi$. This spacetime will play the role of our background. Around it, we introduce perturbations, both in the metric and in the scalar field. We then expand the action in a perturbative series and truncate it at quadratic order, which is the first non-trivial perturbative order. This truncation is essential in all the other steps of the hybrid quantization that we summarize below. For the sake of conciseness, in this work we consider only scalar perturbations. The other physically relevant cosmological perturbations, namely the tensor ones, can be analyzed in a totally parallel way.

The perturbative degrees of freedom can be described with a suitable canonical set of variables, similar to those introduced by Langlois \cite{Langloish}. This set contains the Fourier mode coefficients of the MS gauge invariant perturbations, the Fourier mode coefficients of its momentum (that is also a perturbative gauge invariant), the linear perturbative diffeomorphisms constraints of the system in a convenient Abelianized form, and canonical momenta of the latter that correspond to gauge degrees of freedom. This set is canonical as far as the perturbations are concerned, but it does not include variables for the background. The set can be extended to a canonical one for the whole cosmological system composed by the perturbations and the background along the lines first proposed by Pinto-Neto and his collaborators \cite{PintoNeto,PintoNeto2} and later developed in Ref. \cite{MSCastello}. The result is the inclusion of new canonical variables to describe the background, obtained from the Fourier zero-modes of the metric and the scalar field after correcting them with terms that are quadratic in the perturbations and that respect our order of truncation in the action \cite{MSCastello}. A quantization of this canonical system leads then to physical states that depend only on (a configuration subset of) these background variables and (e.g.) on the real mode coefficients of the MS field. We denote these coefficients by $v_{\vec{k},\varepsilon}$, where $\vec{k}$ is the wavevector of the Fourier mode\footnote{Zero-modes are excluded in the perturbations previous to adopting the continuous (non-compact) limit.} and $\varepsilon$ is a dichotomic label that indicates its parity. 

The hybrid strategy combines a quantum representation of the background (i.e., the corrected zero-modes) with a Fock representation of the MS gauge invariant field. In principle, the quantization of the background variables can be chosen freely except for basic consistency requirements and the need to include a kinematic space of states with an inner product in the FLRW geometry. For the corrected zero-mode of the scalar field and its momentum, we assume a standard representation, with the configuration variable acting by multiplication and its momentum as a derivative. The hybrid quantum system is subject to a global constraint, not yet satisfied in our states, that arises from the zero-mode of the Hamiltonian constraint of the perturbed cosmological model, written in terms of our canonical set of variables. This constraint is the sum of a contribution of the background and another contribution that is quadratic in the perturbations. If we call $\phi$ the (perturbatively corrected) zero-mode of the scalar field, $\pi_{\phi}$ its canonical momentum, $W(\phi)$ the scalar field potential, and $V$ the volume of the compact spatial sections, equal to $8\pi^3$ times the cube of the scale factor, the non-perturbative contribution of the background to the constraint is given in terms of the quantum operators of the chosen representation as $({\hat \pi}_\phi^2- \hat{\mathcal H}_0^{(2)})/2$, with
\begin{equation}
\label{calH_0}
\hat{\mathcal H}_0^{(2)} =\left[\hat{\mathcal H}_0^{(F)}\right]^2-2 W(\phi)\hat{V}^2.
\end{equation}
Here, $\hat{\mathcal H}_0^{(F)}$ can be viewed as a representation of ${\mathcal H}^{(F)}_0= 2\sqrt{3 \pi G}  |\pi_V|V$, where $\pi_V$ would be a momentum variable conjugate to the volume $V$ and $G$ is the Newton constant. We take it as a positive operator by construction. Its square  is the geometric part of the Hamiltonian contribution of the background in the absence of potential and perturbations. On the other hand, for the moment we adopt the operator $2W(\phi)\hat{V}^2$ as a natural choice to represent the potential term. Nonetheless, we leave open the possibility of adopting an alternative representation, taking into account the freedom to add commutators that correspond to different choices of factor orderings (we will actually make use of this freedom in Subsec. \ref{AmodMass}).

An interesting ansatz for the search of physical states consists in a separation of variables in their dependence, so that the wavefunctions factorize in a part that depends on the FLRW geometry and another part that depends on the MS variables. In both partial wavefunctions, a dependence on the scalar field variable $\phi$ is allowed. The partial state for the background, $\chi$, can then be chosen close to a solution of the unperturbed system. Furthermore, as an ingredient of our ansatz, we assume a unitary evolution of this state $\chi$ on $\phi$ that is generated by a positive operator $\hat{\mathcal H}_0$. In consonance with our requirements, and recalling that we are interested in regimes with negligible scalar field potential, we take $\hat{\mathcal H}_0$ as the square root of $\hat{\mathcal H}_0^{(2)}$ (or its positive part \cite{qcCastello}). Therefore, we have
\begin{equation}\label{chievolution}
\chi(V,\phi)=\hat U(V,\phi)\chi_0(V)={\mathcal P}\left[\exp{\left(i\int^{\phi}_{\phi_0} d{\tilde{\phi}}\,\hat{\mathcal H}_0(V,\tilde{\phi})\right)}\right]\chi_0(V),
\end{equation}
where $\chi_0$ is the initial background state at a given $\phi_0$, and the symbol $\mathcal P$ stands for time ordering with respect to $\phi$. 

Introducing this ansatz, and neglecting backreaction and transitions between background states mediated by perturbations,\footnote{One usually neglects as well a typically small term that would obstruct the reality of the MS equations and that, in any case, can be absorbed with a suitable choice of factor ordering in the perturbative contribution to the constraint \cite{MSCastello,qcCastello}.} one can show that the Hamiltonian constraint leads to a dynamics for the MS variables that can be expressed (as a differential equation for the field solutions in our Fock analysis) in the form \cite{Hybridreview,MSCastello,qcCastello}
\begin{equation}\label{modifiedMS}
\ddot{v}_{\vec{k},\varepsilon} + \left[ k^2  + \frac{  \langle \hat{\vartheta}_{e}^q + (\hat{\vartheta}_{o}\hat{{\mathcal H}}_0)_{sym}\rangle_{\chi}}{ \langle \hat{\vartheta}_{e}\rangle_{\chi}} \right] v_{\vec{k},\varepsilon}=0. 
\end{equation}
Here, the dot denotes the derivative with respect to the conformal time, $k$ is the Euclidean norm of the mode wavevector $\vec{k}$, and the subindex $sym$ stands for the symmetrized product. Three new background operators appear in the above formula, that are given by \cite{Hybridreview}
\begin{eqnarray}\label{varthetaEQ}
\hat\vartheta_e^q&=&\frac{1}{2\pi}\widehat{\left[\frac1{V}\right]}^{1/3}\hat{\mathcal H}_0^{(2)}\left(19-18  (\hat{\mathcal H}_0^{(F)})^{-2}\hat{\mathcal H}_0^{(2)}\right) \widehat{\left[\frac1{V}\right]}^{1/3}+\frac{3 }{8 \pi^2 G  }\hat V^{4/3}\left( W^{\prime\prime}(\phi)- \frac{16\pi G}{3} W(\phi)\right),\\ \label{varthetas}
\hat\vartheta_o&=& \frac{3}{\pi}\sqrt{\frac{3}{\pi G}}  W^{\prime}(\phi) \hat V^{2/3} (\hat{\mathcal H}_0^{(F)})^{-1} {\hat\Lambda}_0^{(F)} (\hat{\mathcal H}_0^{(F)})^{-1}\hat V^{2/3} ,\qquad
\hat\vartheta_e=\frac{3 }{2 G}\hat V^{2/3}.
\end{eqnarray}
The prime in the potential denotes the derivative with respect to $\phi$, and the operators $\widehat{[1/V]}$ and ${\hat\Lambda}_0^{(F)}$ have been introduced to take into account some subtleties that arise in the quantization if one follows loop techniques. The operator $\widehat{[1/V]}$  is a regularization of the inverse of the physical volume, and as a result its composition with $\hat{V}$ is not the identity operator for volumes of the Planck order (see e.g. Ref. \cite{MSCastello}). On the other hand, the definition of ${\hat\Lambda}_0^{(F)}$ is related to that of $\hat{\mathcal H}_0^{(F)}$. It represents the quantity $-2\sqrt{3 \pi G} V \pi_V$, so that $|{\hat\Lambda}_0^{(F)}|$ can be viewed as an alternative representation of $\hat{\mathcal H}_0^{(F)}$. In LQC, this alternative representation is obtained using holonomies along edges of the double of the basic coordinate length \cite{MSCastello,Hybridreview}, in order to ensure that the action of the operator leaves invariant the superselection sectors on which the unperturbed Hamiltonian constraint of LQC is defined \cite{MSCastello}. For other quantum representations of the background, these operators may admit simpler definitions that simplify the above expressions. The operators for the inverse and the square inverse of $\hat{\mathcal H}_0^{(F)}$ are well defined because we have assumed in our ansatz that this last operator is positive. As for the conformal time $\eta$ that is used in the modified MS equations \eqref{modifiedMS}, it turns out to be related with the scalar field $\phi$ by \cite{MSCastello,qcCastello}
\begin{equation} \label{conformaltime}
\langle \hat{\cal H}_0\rangle_{\chi} d\eta = \langle\hat{ \vartheta}_e\rangle_{\chi}  d\phi.
\end{equation}

The modified MS equation \eqref{modifiedMS} can be written in the form $\ddot{v}_{\vec{k},\varepsilon} + [ k^2  + s(\eta)]v_{\vec{k},\varepsilon} =0$, with an effective time-dependent mass $s$ equal to a ratio of expectation values. According to the arguments of Ref. \cite{BeaDiagonal2}, that are based in turn on the criterion of asymptotic Hamiltonian diagonalization for the choice of vacuum state put forward in Ref. \cite{BeaDiagonal}, a preferred set of (normalized) positive frequency solutions $\{\mu_k\}$ of that equation is 
\begin{equation}\label{muh}
\mu_k=\frac{1}{\sqrt{-2\text{Im}(h_k)}}e^{i\int^{\eta}_{\eta_0} d\tilde{\eta} \, \text{Im}(h_k)},
\end{equation}
where $\text{Im}$ is the imaginary part, $\eta_0$ is any fixed reference time, and $h_k$ is a solution of a Riccati equation that contains the effective mass, namely
\begin{equation}\label{hache}
\dot{h}_{k}=k^2+s+h_k^2.
\end{equation}
More specifically, the desired solution must admit the following asymptotic expansion for infinitely large wavenumbers \cite{BeaDiagonal2}:
\begin{equation}\label{asymptotich} 
\frac{1}{h_k}\sim  \frac{i}{k}\left[1-\frac{1}{2k^2}\sum_{n=0}^{\infty}\left(\frac{-i}{2k}\right)^{n}\gamma_n \right],
\end{equation}
with $\gamma_{0}=s$ and the rest of gamma-coefficients determined by the recurrence relation
\begin{equation}\label{recursion}
\gamma_{n+1}=-\dot{\gamma}_{n}+4s \left[\gamma_{n-1}+\sum_{m=0}^{n-3}\gamma_m \gamma_{n-(m+3)}\right]-\sum_{m=0}^{n-1}\gamma_m \gamma_{n-(m+1)}.
\end{equation}
Notice that the effective mass fixes $\gamma_0$ and also appears explicitly in this relation. 

If the time interval where the positive frequency solutions $\{\mu_k\}$ are valid includes the end of inflation, the power spectrum of the scalar perturbations is given, up to a $k$-independent multiplicative factor, by \cite{Langlois}
\begin{equation}\label{powerspectrum}
\mathcal{P}_{\mathcal{V}}(k,\eta)=\frac{k^3}{2\pi^2}|\mu_k (\eta)|^2=-\frac{k^3}{4\pi^2 \text{Im}(h_k)(\eta)},
\end{equation}
evaluated at a time $\eta$ when slow-roll inflation has just finished. If the aforementioned interval does not cover the inflationary period, one can use our solutions to obtain initial values for the primordial perturbations at the end of the kinetically dominated regime and continue the evolution afterwards with those values.

\section{Leading-order corrections to the effective mass}\label{modMass}

Since we are interested in studying kinetically dominated regimes, in this section we want to calculate the leading-order correction of the potential to the effective mass of the perturbations around the free scalar field case. For concreteness, we consider the expression of the effective mass $s$ derived in hybrid quantum cosmology \cite{MSCastello}, which, according to our discussion in the previous section, is given by the following ratio of expectation values:
\begin{equation}\label{modifiedMass}
s=  \frac{  \langle \hat{\vartheta}_{e}^q + (\hat{\vartheta}_{o}\hat{{\mathcal H}}_0)_{sym}\rangle_{\chi}}{ \langle \hat{\vartheta}_{e}\rangle_{\chi}} .
\end{equation}
For cosmological scenarios in General Relativity, one can take the classical limit of the above expression, replacing operators by functions on the classical phase space and expectation values by evaluation on classical solutions, after having taken due care of the detailed effects of the potential. We divide our calculations in this section in three parts. First, we consider the influence of the potential at leading order in $W(\phi)$ and its derivatives on the relevant operators and in the evolution of the background state $\chi$. Since this evolution is usually described in terms of the scalar field $\phi$, we then discuss the leading-order effect of the potential in the change to conformal time, which is the parameter used in the modified MS equations for the perturbations. Finally, we combine these two classes of corrections to obtain the leading-order effect of the potential in the effective mass regarded as a function of the conformal time, $s=s(\eta)$. We also succinctly consider the particularization of our discussion to some classical and quantum cosmology models.

\subsection{Corrections to the effective mass in terms of the scalar field}\label{AmodMass}

In the following, by linear or higher-order dependence in the potential we understand a dependence on $W(\phi)$ or any of its derivatives (or products of them). The leading-order contribution of the potential to the different operators that appear explicitly in the ratio of expectation values that determines the effective mass is easy to calculate. First, we notice that $\hat{\vartheta}_e$, appearing in the denominator, is independent of the potential. On the other hand, $\hat{\vartheta}_o$ is linear in the potential, without higher-order corrections. Then, its symmetrized product with $\hat{\mathcal{H}}_0$ has a linear contribution of the potential that, with our previous definitions, is simply given by the symmetrized product with $\hat{\mathcal{H}}_0^{(F)}$. Recall that this last operator is just the generator of the background evolution in the free case, and was introduced in Eq. \eqref{calH_0}. Finally, in the case of $\hat{\vartheta}_e^q$, it is straightforward to see from our definitions that
\begin{equation}\label{varthetaEQlinear}
\hat{\vartheta}_e^q = \frac{1}{2\pi} \widehat{\left[ \frac{1}{V}\right]}^{1/3} \left( \hat{\mathcal{H}}_0^{(F)}\right)^2  \widehat{\left[ \frac{1}{V}\right]}^{1/3}+ \left\{\frac{17}{\pi} W(\phi) \widehat{\left[ \frac{1}{V}\right]}^{1/3} \hat{V}^2 \widehat{\left[ \frac{1}{V}\right]}^{1/3} + \left( \frac{3}{8\pi^2 G} W''(\phi) - \frac{2}{\pi} W(\phi)\right) \hat{V}^{4/3}\right\}+O(W^2),
\end{equation}
where the first term is the free contribution, the term in curved brackets is the leading-order contribution of the potential, and the terms $O(W^2)$ are quadratic or higher-order  in the potential.

The background state $\chi$ also introduces a dependence on the potential because its evolution from the initial state $\chi_0(V)$ depends on it. In this subsection, we calculate the leading-order dependence when the evolution is described in terms of the scalar field $\phi$. This dependence can be extracted using an interaction picture for the quantum evolution, along the lines explained in Ref. \cite{qcCastello}. However, employing the freedom in the choice of factor ordering of the contribution of the potential that we commented in Eq. \eqref{calH_0}, we use a different approximation to the generator $\hat{\mathcal{H}}_0$ around the free Hamiltonian $\hat{\mathcal{H}}_0^{(F)}$, which proves to be more convenient for the rest of our calculations. Since most of the arguments parallel the discussion of Ref. \cite{qcCastello}, we only explain here the most important steps. 

We first want to approximate $\hat{\mathcal{H}}_0$ around the generator of the free evolution up to quadratic and higher-order terms in the potential. From the definition of $\hat{\mathcal{H}}_0$ as the square root of $\hat{\mathcal{H}}_0^{(2)}$, it is not difficult to check that our operator can be represented in the form
\begin{equation}\label{WH_0}
\hat{\mathcal{H}}_0=\hat{\mathcal{H}}_0^{(F)}- W(\phi) (\hat{\mathcal{H}}_0^{(F)})^{-1/2} \hat{V}^2 (\hat{\mathcal{H}}_0^{(F)})^{-1/2} + O(W^2).
\end{equation}
Our statement can be checked by taking its square and realizing that the result gives indeed $\hat{\mathcal{H}}_0^{(2)}$ at the considered order in the potential, up to double commutators that can be absorbed with a suitable factor ordering (see Ref. \cite{qcCastello} and our remarks after Eq. \eqref{calH_0}). The second term on the right-hand side of our equation then provides the leading-order contribution of the potential.

Let us call $\chi^{(F)}(V,\phi)$ the background state that would result from the initial state $\chi_0(V)$ if the evolution in $\phi$ were generated by the Hamiltonian of the free case. Explicitly,
\begin{equation}\label{chiF}
\chi^{(F)}(V,\phi)= \exp{[i \hat{\mathcal{H}}_0^{(F)}(\phi-\phi_0)]} \chi_0(V).
\end{equation}
Employing the interaction picture in a convenient way, it is possible to show that
\begin{equation}\label{chi-chiF}
\chi(V,\phi)= \exp{[i \hat{\mathcal{H}}_0^{(F)}(\phi-\phi_0)]} {\mathcal P}\left[\exp{\left(i\int^{\phi}_{\phi_0} d{\tilde{\phi}}\,\hat{\mathcal H}_1^{(I)}(V,\tilde{\phi})\right)}\right]\exp{[-i \hat{\mathcal{H}}_0^{(F)}(\phi-\phi_0)]} \chi^{(F)}(V,\phi),
\end{equation} 
where we have defined 
\begin{equation}\label{H1I}
\hat{\mathcal H}_1^{(I)}= \exp{[-i \hat{\mathcal{H}}_0^{(F)}(\phi-\phi_0)]} \left(\hat{\mathcal{H}}_0-\hat{\mathcal{H}}_0^{(F)}\right) \exp{[i \hat{\mathcal{H}}_0^{(F)}(\phi-\phi_0)]}.
\end{equation}
The two last expressions are exact to all orders in the potential. Using them and our approximation \eqref{WH_0}, it is now easy to obtain the leading-order correction of the potential to the freely evolved background state. With the notation $\chi(V,\phi)=\hat{U}^{(W)}(\phi) \chi^{(F)}(V,\phi)$, we get
\begin{eqnarray}\label{UW}
\hat{U}^{(W)} (\phi)&=&1 - i \int^{\phi}_{\phi_0}d\tilde{\phi} \, \hat{K}(\phi,\tilde{\phi}) W(\tilde{\phi}) + O(W^2), \\
\label{Kphi}
\qquad \hat{K}(\phi,\tilde{\phi})&=& \exp{[i \hat{\mathcal{H}}_0^{(F)}(\phi-\tilde{\phi})]}(\hat{\mathcal{H}}_0^{(F)})^{-1/2} \hat{V}^2 (\hat{\mathcal{H}}_0^{(F)})^{-1/2}\exp{[-i \hat{\mathcal{H}}_0^{(F)}(\phi-\tilde{\phi})]}.
\end{eqnarray}
Notice that, with our choice of factor ordering in Eq. \eqref{WH_0}, the two operators on the right of $ \hat{V}^2$  commute, as well as the two operators in front of it. Furthermore, the operator $\hat{K}(\phi,\tilde{\phi})$ is symmetric since so are $\hat{\mathcal{H}}_0^{(F)}$ and $\hat{V}$, by construction. The above expressions allow us to pass from the expectation values on the exact state $\chi(V,\phi)$ that determine the effective mass to expectation values on $\chi^{(F)}(V,\phi)$, which evolves in the scalar field $\phi$ according to the free dynamics. We notice the different approach taken here with respect to the discussion presented in Ref. \cite{qcCastello}. Here, the expectation values are finally referred to the case of the free evolution, which is left unspecified, while in Ref. \cite{qcCastello} these expectation values were directly rewritten in terms of the initial background state $\chi_0(V)$.

\subsection{Corrections in the relation between the scalar field and the conformal time}\label{BmodMass}

In order to obtain the effective mass in terms of the conformal time, we still have to invert the relation between this time and the scalar field used for the parametrization of the quantum background dynamics, according to Eq. \eqref{conformaltime}. Again, we are interested in extracting the relation for the free case and the leading-order correction to it produced by the presence of a potential. One can compute them using the expressions for the operator $\hat{\mathcal{H}}_0$ and for the background state $\chi(V,\phi)$ at leading order in the potential, provided by Eqs. \eqref{WH_0} and \eqref{UW} respectively. A simple calculation at this order leads then to the relation
\begin{eqnarray}\label{modifiedconformaltime}
&& \frac{3}{2G}\langle \hat{V}^{2/3} \rangle_{\chi^{(F)}}\,  d\phi -i \frac{3}{2G}\langle  I_W^{(F)}[\hat{V}^{2/3}]  \rangle_{\chi^{(F)}}\, d\phi = \langle \hat{\mathcal{H}}_0^{(F)}  \rangle_{\chi^{(F)}}\, d\eta \nonumber \\
&&- W(\phi)    \langle  (\hat{\mathcal{H}}_0^{(F)})^{-1/2} \hat{V}^2 (\hat{\mathcal{H}}_0^{(F)})^{-1/2}  \rangle_{\chi^{(F)}}\, d\eta   -i  \langle    I_W^{(F)}[ \hat{\mathcal{H}}_0^{(F)} ]        \rangle_{\chi^{(F)}}\, d\eta,
\end{eqnarray}
where we have introduced the notation
\begin{equation}\label{IWF}
I_W^{(F)}[\hat{A}]= \int^{\phi}_{\phi_0}d\tilde{\phi} \, W(\tilde{\phi} ) [\hat{A}, \hat{K}(\phi,\tilde{\phi})] ,
\end{equation}
for any operator $\hat{A}$. In the case of $\hat{\mathcal{H}}_0^{(F)}$, we notice that this operator commutes with all the factors of $\hat{K}(\phi,\tilde{\phi})$ except with the central one, $\hat{V}^2$. 

Neglecting in our expressions the contribution of the potential, we obtain the relation for the free evolution, that we call $\eta^{(F)}(\phi)$. Explicitly, 
\begin{equation}\label{etaF}
\eta^{(F)}(\phi) = \frac{3}{2G} \int_{\phi_0}^{\phi}  d\tilde{\phi}\, \frac{\langle \hat{V}^{2/3} \rangle_{\chi^{(F)}} }{\langle \hat{\mathcal{H}}_0^{(F)} \rangle_{\chi^{(F)}} },
\end{equation}
where we have set, without loss of generality, $\eta^{(F)}(\phi_0)=0$, and the expectation values in the integrand are computed on the free evolved state $\chi^{(F)}(V,\tilde{\phi})$. This is why the integral cannot be calculated trivially and needs the knowledge of (only) the free evolution (but not of the dynamics when the potential is present).

This relation can now be substituted in the terms that are linear in the potential in Eq. \eqref{modifiedconformaltime} preserving our leading-order approximation. Thus, in the last two terms of that equation, we can substitute $d\eta$ by $(\eta^{(F)})^{\prime}d\phi$ (the prime denoting the derivative with respect to $\phi$). Adopting the notation $\eta(\phi)=\eta^{(F)}(\phi)+\eta^{(W)}(\phi)+O(W^2)$ for the corrections introduced in the dependence of the conformal time by the presence of a potential, we conclude then from Eqs. \eqref{modifiedconformaltime} and \eqref{etaF} that
\begin{eqnarray}\label{etaW}
\eta^{(W)}(\phi) &=& \frac{3}{2G} \int^{\phi}_{\phi_0}   d\tilde{\phi}\,   \frac{1} { \big[\langle \hat{\mathcal{H}}_0^{(F)} \rangle_{\chi^{(F)}}\big]^{2}}  \Big[ \left( W(\tilde{\phi})    \langle  (\hat{\mathcal{H}}_0^{(F)})^{-1/2} \hat{V}^2 (\hat{\mathcal{H}}_0^{(F)})^{-1/2}  \rangle_{\chi^{(F)}} + i  \langle    I_W^{(F)}[ \hat{\mathcal{H}}_0^{(F)} ]        \rangle_{\chi^{(F)}}\right) \langle \hat{V}^{2/3} \rangle_{\chi^{(F)}}   
\nonumber \\
& - &i \langle \hat{\mathcal{H}}_0^{(F)} \rangle_{\chi^{(F)}}  \langle  I_W^{(F)}[\hat{V}^{2/3}]  \rangle_{\chi^{(F)}}  \Big]  .
\end{eqnarray}

Actually, in order to obtain the effective mass in terms of the conformal time, we need the inverse of the relation that we have computed, inverse which determines the dependence of the scalar field $\phi$ as a function of $\eta$. Let us call $\phi^{(F)}(\eta)$ the inverse of the functional dependence \eqref{etaF} for the free evolution [so that $\phi^{(F)}(\eta^{(F)}(\phi))=\phi$ and $\eta^{(F)}(\phi^{(F)}(\eta))=\eta$]. Then, at our leading-order approximation, we can express the relation that we are looking for in the form $\phi(\eta)=\phi^{(F)}(\eta)+\phi^{(W)}(\eta)+O(W^2)$ with the following leading-order correction of the potential to the conformal time:
\begin{equation}\label{phiW}
\phi^{(W)}(\eta)=-\phi^{(F)}\left(\eta^{(W)}\big(\phi^{(F)}(\eta)\big)\right).
\end{equation}
This is minus the composition of the inverse of the free conformal time function with the correction produced by the potential and composed again with the inverse of the free conformal time function.

\subsection{Corrections to the effective mass in terms of the conformal time. Applications}\label{CmodMass}

We can finally derive the leading-order correction to the effective mass, treated as a function of the conformal time. First, let us define
\begin{equation}\label{chiFeta}
\tilde{\chi}^{(F)}(V,\eta)= \exp{\big[i \hat{\mathcal{H}}_0^{(F)}\big(\phi^{(F)}(\eta)-\phi_0\big)\big]} \chi_0(V).
\end{equation}
According to our discussion above, we have that 
\begin{equation}\label{chiFtot}
\chi^{(F)}\big(V,\phi(\eta)\big)= \exp{\big[i \hat{\mathcal{H}}_0^{(F)}\big(\phi(\eta)-\phi^{(F)}(\eta)\big)\big]} \tilde{\chi}^{(F)}(V,\eta).
\end{equation}
This expression is exact. Approximating it to leading order in the potential, we obtain
\begin{equation}\label{chiFtotapp}
\chi^{(F)}\big(V,\phi(\eta)\big)= \left[1+i \hat{\mathcal{H}}_0^{(F)} \phi^{(W)}(\eta) +O(W^2) \right] \tilde{\chi}^{(F)}(V,\eta),
\end{equation}
with $\phi^{(W)}(\eta)$ given in Eq. \eqref{phiW}.  Combining this with the corrections of the background state already derived in terms of the conformal time [see Eq. \eqref{UW}], we conclude that
\begin{equation}\label{chitot}
\chi\big(V,\phi(\eta)\big)= \left[1+i \hat{\mathcal{H}}_0^{(F)} \phi^{(W)}(\eta) - i \int^{\phi^{(F)}(\eta)}_{\phi_0}d\tilde{\phi} \, \hat{K}\big(\phi^{(F)}(\eta),\tilde{\phi}\big) W(\tilde{\phi}) + O(W^2)\right] \tilde{\chi}^{(F)}(V,\eta).
\end{equation}
We recall that the operator $\hat{K}(\phi,\tilde{\phi})$ was defined in Eq. \eqref{Kphi}. 

Other than via their dependence on the state $\chi$, the dependence on $\phi(\eta)$ of the expectation values that provide the effective mass appears only in the leading-order correction of the potential, but not on the free contribution independent of it. Therefore, at the dominant order that we are considering, the expression of those leading-order corrections in terms of the conformal time can be obtained by simply evaluating the scalar field on the trajectory of the free case, namely on $\phi^{(F)}(\eta)$. 

With these results, it is then easy to check that the factor corresponding to the denominator of the effective mass \eqref{modifiedMass} in conformal time can be expressed at leading order in the potential as
\begin{equation}\label{varthetaEconf}
\frac{1}{\langle\hat{\vartheta}_e \rangle_{\chi}} =\frac{2G}{3 \langle \hat{V}^{2/3} \rangle_{\tilde{\chi}^{(F)}} } \left[ 1 - 
i\frac{\langle  [\hat{V}^{2/3},\hat{J}^{(W)} ] \rangle_{\tilde{\chi}^{(F)}} }{\langle \hat{V}^{2/3} \rangle_{\tilde{\chi}^{(F)}}}      \right]+O(W^2),
\end{equation}
where, to shorten our notation, we have called 
\begin{equation}\label{JW}
\hat{J}^{(W)}(\eta)=\hat{\mathcal{H}}_0^{(F)} \phi^{(W)}(\eta) -  \int^{\phi^{(F)}(\eta)}_{\phi_0}d\tilde{\phi} \, \hat{K}\big(\phi^{(F)}(\eta),\tilde{\phi}\big) W(\tilde{\phi}).
\end{equation} 

One can similarly approximate the numerator of the effective mass, expressed in conformal time, employing Eq. \eqref{varthetaEQlinear} for the correction of $\hat{\vartheta}^q_e$ and recalling that the other summand can be replaced at leading order in the potential with $(\hat{\vartheta}_o\hat{\mathcal{H}}_0^{(F)})_{sym}$. Combining this with Eq. \eqref{varthetaEconf}, we obtain that
\begin{eqnarray}\label{stot}
s(\eta)&=&s^{(F)}(\eta)+s^{(W)}(\eta)+O(W^2),\\  \label{sF}
s^{(F)}(\eta) &=&  \frac{G}{3\pi}  \frac{\langle \widehat{\left[\frac{1}{V}\right]}^{1/3}  ( \hat{\mathcal{H}}_0^{(F)})^2 \widehat{\left[\frac{1}{V}\right]}^{1/3} \rangle_{\tilde{\chi}^{(F)}}} {\langle \hat{V}^{2/3} \rangle_{\tilde{\chi}^{(F)}}}, 
\end{eqnarray}
with the leading-order correction given by
\begin{eqnarray} \label{sW}
s^{(W)}(\eta)&=&   
\frac{2G}{3\pi\langle \hat{V}^{2/3} \rangle_{\tilde{\chi}^{(F)}}}  \Bigg\{\left( \frac{3}{8\pi G} W'' - 2 W\right)  \langle \hat{V}^{4/3} \rangle_{\tilde{\chi}^{(F)}}  +  17 W \langle \widehat{\left[ \frac{1}{V}\right]}^{1/3} \hat{V}^2 \widehat{\left[ \frac{1}{V}\right]}^{1/3} \rangle_{\tilde{\chi}^{(F)}}  \Bigg\} \nonumber \\
&+&\frac{2\sqrt{3G}}{\pi \sqrt{\pi}\langle \hat{V}^{2/3} \rangle_{\tilde{\chi}^{(F)}}}  W^{\prime}  \langle (  \hat V^{2/3} (\hat{\mathcal H}_0^{(F)})^{-1} {\hat\Lambda}_0^{(F)} (\hat{\mathcal H}_0^{(F)})^{-1}\hat V^{2/3} \hat{\mathcal{H}}_0^{(F)})_{sym}\rangle_{\tilde{\chi}^{(F)}}    \nonumber\\
&+& \frac{i} {\langle \hat{V}^{2/3} \rangle_{\tilde{\chi}^{(F)}} } \Bigg\{- s^{(F)} \langle [\hat{V}^{2/3},\hat{J}^{(W)}] \rangle_{\tilde{\chi}^{(F)}}  
 + \frac{G}{3\pi} \langle  \Big[ \widehat{\left[\frac{1}{V}\right]}^{1/3} ( \hat{\mathcal{H}}_0^{(F)})^2  \widehat{\left[\frac{1}{V}\right]}^{1/3}, \hat{J}^{(W)}\Big] \rangle_{\tilde{\chi}^{(F)}} \Bigg\}.
\end{eqnarray}
In the last formula, the potential and its derivatives are evaluated at $\phi^{(F)}(\eta)$, and the mass $s^{(F)}$, the state $\tilde{\chi}^{(F)}$, and the operator $\hat{J}^{(W)}$ at $\eta$.

The application of our formulas for the free effective mass and its leading-order correction in the case of classical General Relativity can be done as follows. First of all, we notice that the last line of Eq. \eqref{sW} is the implicit correction of the potential to the purely free contribution $s^{(F)}(\eta)$ of the mass. Then, passing from operators to functions on phase space, ignoring the distinction between $|\hat{\Lambda}_0^{F}|$ and $\hat{\mathcal{H}}_0^{(F)}$, identifying the classical analog of the inverse volume operator as $1/V$ in this passage, translating the commutators of operators in the last line of Eq. \eqref{sW} into $i$ times the corresponding Poisson brackets $\{ \, , \, \}$, and  interpreting the expectation values as the evaluation on free classical solutions, we get
\begin{eqnarray}\label{sFGR}
s^{(F)}_{GR}&=&\frac{G}{3\pi} \frac{({\mathcal{H}}_0^{(F)})^2}{V^{4/3}}=4 G^2 V^{2/3} \pi_V^2, \\ \label{sWGR}
s^{(W)}_{GR}&= &\frac{2G}{3\pi} V^{2/3} \left(\frac{3}{8\pi G}W^{\prime\prime}+15 W -3\sqrt{\frac{3}{\pi G}} \, {sgn}{(\pi_V)} W^{\prime}\right) - 4G^2\{V^{2/3} \pi_V^2,J^{(W)}\}.
\end{eqnarray}
All of the phase space functions that appear in these two formulas must be evaluated on free classical solutions. The last term in the second equation is the leading-order correction of the potential coming from the evaluation of $V^{2/3} \pi_V^2$ on exact classical solutions instead of free ones, and it is the contribution of the last line in Eq. \eqref{sW}. In our calculations, we have used the identification ${\mathcal{H}}_0^{(F)}=2\sqrt{3\pi G} |\pi_V| V$, the symbol ${sgn}$ denotes the sign function, and $J^{(W)}$ is the classical counterpart of the operator $\hat{J}^{(W)}$, namely
\begin{equation}\label{JWGR}
J^{(W)}= 2\sqrt{3\pi G} |\pi_V| V \phi^{(W)} - \frac{1}{2\sqrt{3\pi G} |\pi_V| V } \int ^{\phi^{(F)}}_{\phi_0}d\tilde{\phi}\, W(\tilde{\phi})\sum_{n=0}^{\infty}\frac{1}{n!}\big(\phi^{(F)}-\tilde{\phi}\big)^{n}\left\lbrace V^2,{\mathcal{H}}_0^{(F)}\right\rbrace_{(n)},
\end{equation}
where we have obviated the explicit dependence of $\phi^{(F)}$ and $\phi^{(W)}$ on the conformal time $\eta$, and $\{ V^2 , {\mathcal{H}}_0^{(F)} \}_{(n)}$ is the $n$th-order Poisson bracket of $V^2$ with the generator ${\mathcal{H}}_0^{(F)}$ of the free evolution. These brackets appear because the classical counterpart of Eq. \eqref{Kphi} is the free evolution of $V^2/{\mathcal{H}}_0^{(F)}$ from $\tilde{\phi}$ to $\phi^{(F)}$ (and ${\mathcal{H}}_0^{(F)}$ remains constant along this free evolution). In particular, we notice that, when the potential is set to zero, the only non-vanishing component $s^{(F)}_{GR}$ of the mass turns out to reproduce the standard function $-\ddot{z}/z=-\ddot{a}/a$ in General Relativity, with $z=a^2 \dot{\phi}/\dot{a}$.

The operator representation for the case of hybrid LQC is given in Sec. 6 of Ref. \cite{MSCastello} and we do not repeat it here. In that reference, $W(\phi)$ was particularized to a quadratic potential, $m^2 \phi^2/2$. All relevant operators for our calculations can be constructed in terms of the volume operator, which acts by multiplication in the volume representation adopted in Ref. \cite{MSCastello}, and of an operator $\hat{\Omega}_0$ defined by means of holonomies, which shifts the eigenvalues of the volume eigenstates in a way that depends on the so-called Immirzi parameter of Loop Quantum Gravity \cite{Immirzi}. For the specific definition that we adopt here for $\hat{\Omega}_0$ and the conventions about the numerical factors in this definition, we use Eq. (28) of Ref. \cite{Hybridreview}. In particular, in this manner we have $\hat{\mathcal{H}}_0^{(F)}=|\hat{\Omega}_0| $. The operator $\hat{\Lambda}_0^{F}$ is defined in terms of holonomies exactly in the same way as $\hat{\Omega}_0$, except for the fact that the coordinate length of the edges of the holonomies are doubled. Finally, the inverse volume operator is obtained by means of a regularization that uses holonomies. It is a standard operator in LQC, and its action on volume eigenstates can be found, e.g., in Ref. \cite{MSCastello}.

Finally, let us briefly comment on the case of a hybrid quantization based on geometrodynamics \cite{QCHalliwell,LQC1}. We can start with operators $\hat{a}$ and $\hat{\pi}_a$ for the scale factor and its momentum, defined in a geometrodynamic representation in which they respectively act by multiplication, $\hat{a}=a$, and by differentiation. The representation is provided with an inner product for the background geometry that is given by the integration over the scale factor with a certain continuous measure (not necessarily $da$). The volume and inverse volume operators can be defined, respectively, as $8 \pi^3 a^3$ and $1/(8 \pi^3 a^3)$. We can also define $\hat{\Lambda}_0^{(F)}= -(a \hat{\pi}_a+\hat{\pi}_a a) \sqrt{\pi G/3}$, and $\hat{\mathcal{H}}_0^{(F)}$ exactly in the same way but replacing $-\hat{\pi}_a$ with $|\hat{\pi}_a|$. This simplifies some products of operators in our expressions for the free mass $s^{(F)}$ and its leading-order corrections, for instance the product of powers of the inverse volume and of the volume in the second summand of $s^{(W)}$ in Eq. \eqref{sW}.  With all these guidelines, it is a simple exercise to particularize our results to hybrid geometrodynamic quantum cosmology.

\section{Leading-order corrections in the positive frequency solutions}\label{modFreq}

Although we have discussed the corrections that the presence of a potential introduces at leading order in the effective mass that appears in the MS equations, and therefore how it affects the propagation of the scalar perturbations, in order to integrate these equations we still need suitable initial conditions. In a conventional inflationary period that can be treated as an approximate epoch of de Sitter inflation, a natural choice of initial conditions are those corresponding to a Bunch-Davies state \cite{BunchDavies}. Nonetheless, in scenarios with kinetically dominated regimes, and even with bounces originated by quantum phenomena, the Bunch-Davies state does not seem to be an adequate choice of state  any longer, at least for wavelengths comparable to the natural scales involved in these scenarios, e.g. the Hubble radius at the end of the fast-roll period or the typical scale of the possible quantum phenomena. As we have commented in Sec. \ref{modMass}, in these circumstances an alternative appealing choice of vacuum state, and therefore of initial conditions, has recently been put forward, based on a diagonalization of the contribution of the perturbations to the Hamiltonian constraint which removes the production of particle pairs, at least asymptotically in the sector of modes with large wavenumber $k$ \cite{BeaDiagonal,BeaDiagonal2}. Moreover, this choice of state seems to lead to power spectra without superimposed rapid oscillations  with respect to the wavenumber that, when averaged over resolution bins, would artificially increase the predicted power \cite{Bealast,BeaDiagonal2}. 

This vacuum state with good properties is associated with the choice of a set of positive frequency solutions that are determined by functions of the conformal time, $h_k$, which are solutions of the Riccati equation \eqref{hache}. Notice that this equation contains the effective mass $s$. The specific solution $h_k$ of interest has to admit an asymptotic expansion of the form \eqref{asymptotich}, with coefficients $\gamma_n$ that are independent of $k$ and satisfy the recursive relation \eqref{recursion}. This relation contains again the effective mass $s$. In addition, the starting datum to solve it is the identification $\gamma_0=s$. Thus, we see that the modification of the effective mass away from the free-evolution value $s^{(F)}$ introduces changes in the equation that $h_k$ has to solve and in its characteristic asymptotic expansion. In this section, we will discuss the leading-order corrections introduced by the scalar field potential $W(\phi)$ in $h_k$ and the associated asymptotic coefficients $\gamma_n$.  

Similar to the approximations that we have carried out in the previous section, we take $h_k=h_k^{(F)}+h_k^{(W)}+O(W^2)$. The function $h_k^{(F)}$ is a solution to the Riccati equation \eqref{hache} with the effective mass $s$ replaced with the mass $s^{(F)}$ for a free scalar field. In addition, $h_k^{(F)}$ admits an asymptotic expansion of the form \eqref{asymptotich} with coefficients $\gamma_n^{(F)}$ that satisfy the recursive relation obtained with the replacement of the effective mass with $s^{(F)}$. Moreover, the recursion starts with the identification $\gamma_0^{(F)}=s^{(F)}$. In other words, $h_k^{(F)}$ is indeed the solution for an asymptotic diagonalization of the Hamiltonian contribution of the perturbations in absence of a scalar field potential. Assuming that we can determine $h_k^{(F)}$ completely, the question that we want to investigate is whether we can then fix $h_k^{(W)}$ in terms of the potential, respecting the studied leading order in all our considerations.

One can easily check that the corrections that the potential introduces in the Riccati equation with respect to the free case lead, at this leading order, to the following linear, first-order, ordinary differential equation:
\begin{equation}\label{EDOhache}
\dot{h}_k^{(W)}(\eta) = s^{(W)}(\eta) +2 h_k^{(F)}(\eta) h_k^{(W)}(\eta). 
\end{equation}
This equation can be integrated exactly, provided that $h_k^{(F)}$ and $s^{(W)}$ are known. The general solution is
\begin{equation}\label{hW}
h_k^{(W)} (\eta) = \left( C_k^{(W)} + \int_{\eta_0}^\eta d\tilde{\eta} \, s^{(W)}(\tilde{\eta}) e^{-I_{h_k}^{(F)}(\tilde{\eta})}  \right) e^{I_{h_k}^{(F)}(\eta)}, 
\end{equation}
where $C_k^{(W)}$ is an integration constant that we allow to depend on the particular potential $W$ under consideration (or its derivatives), $\eta_0$ is an arbitrary initial time, and we have called
\begin{equation}\label{IhkF}
I_{h_k}^{(F)}(\eta) = 2 \int_{\eta_0}^\eta d\tilde{\eta}\, h_k^{(F)}(\tilde{\eta}).
\end{equation}
Moreover, the demand that $h_k^{(W)} $ must vanish when the potential is not present, and hence when $s^{(W)}$ is zero, requires that the constant $C_k^{(W)}$ vanish when the potential is identically zero (for all wavenumbers $k$). Apart from this requirement, $C_k^{(W)}$ remains unspecified for the time being. If we integrate in our solution the integral that contains $s^{(W)}$ iteratively by parts, we can formally express our function $h_k^{(W)}$ as
\begin{equation}\label{hkWbyparts}
h_k^{(W)}(\eta)= - \sum_{n=0}^{\infty} \frac{d^n}{d \tau_k^n} \left(\frac{s^{(W)}(\eta)}{2h_k^{(F)}(\eta)} \right)+ D_k^{(W)}(\eta_0) e^{I_{h_k}^{(F)}(\eta)},
\end{equation}
where we have defined a mathematically convenient time $d\tau_k=2h_k^{(F)}(\eta) d\eta$, and $D_k^{(W)}(\eta_0)$ is a constant that depends on $\eta_0$ and $C_k^{(W)}$, given by
\begin{equation}\label{Dk}
D_k^{(W)} (\eta_0)= \sum_{n=0}^{\infty} \left. \frac{d^n}{d \tau_k^n} \left(\frac{s^{(W)}(\eta)}{2h_k^{(F)}(\eta)} \right)\right\vert_{\eta=\eta_0} + C_k^{(W)}.
\end{equation}

On the other hand, the fact that $h_k$ must admit a specific asymptotic expansion fixed by the effective mass $s$, while its free scalar field contribution $h_k^{(F)}$ satisfies a similar asymptotic expansion fixed by the effective mass of the free case $s^{(F)}$, imposes a restriction on the asymptotic behavior of the leading-order correction $h_k^{(W)}$. Dividing $h_k^{(F)}$ by $h_k$, using that the latter is equal to $h_k^{(F)}+h_k^{(W)}$ up to corrections of order $O(W^2)$, and employing the asymptotic expansion \eqref{asymptotich} and its counterpart for $h_k^{(F)}$, we obtain that, in the asymptotic limit of infinitely large $k$, 
\begin{equation}\label{asymptoticratio}
 \frac{h_k^{(F)}}{h_k} = 1 - \frac{h_k^{(W)}}{h_k^{(F)}} + \mathcal{O}(W^2) \sim \frac{ 1 +2 \sum_{n=0}^\infty ( 2ik)^{-n-2}  \gamma_n   }{   1 +2  \sum_{n=0}^\infty ( 2ik )^{-n-2} \gamma_n^{(F)} }.
 \end{equation}
Let us adopt an expansion for the coefficients $\gamma_n$ similar to the one that we have introduced for $h_k$ and other quantities, expressing these coefficients as those corresponding to the case of a free scalar field, $\gamma_n^{(F)}$, plus a leading-order correction linear in the the potential, $\gamma_n^{(W)}$, plus other higher-order corrections, namely, $\gamma_n=\gamma_n^{(F)}+\gamma_n^{(W)}+O(W^2)$. It then follows from Eq. \eqref{asymptoticratio} (using again the asymptotic expansion of $h_k^{(F)}$) that, at the considered order in the potential, 
 \begin{equation}\label{asymptoticW} 
h_k^{(W)} \sim  \frac{ \sum_{n=0}^\infty ( 2ik )^{-n-1} \gamma_n^{(W)} }{  \left[ 1 + 2 \sum_{n=0}^\infty ( 2ik)^{-n-2} \gamma_n^{(F)}\right]^2}.
 \end{equation}
The right-hand side of this asymptotic equality can be expanded as a series in inverse powers of $k$ in the limit $k\rightarrow\infty$. The coefficients $\gamma_n^{(W)}$ satisfy a recurrence relation that can be obtained by substracting to Eq. \eqref{recursion} its analog for the free case and keeping in the result only terms that have a linear functional dependence on the potential. In this way, one obtains the following relation at our approximation order
\begin{equation}\label{gammaW}
\gamma_{n+1}^{(W)} = -\dot{\gamma}_n^{ (W)}  + 4s^{(F)} \left[\gamma_{n-1}^{(W)}  + 2\sum_{m=0}^{n-3} \gamma_m^{(F)}  \gamma_{n-(m+3)}^{(W)}  \right] - 2\sum_{m=0}^{n-1} \gamma_m^{(F)} \gamma^{(W)} _{n-(m+1)} + 4s^{(W)} \left[\gamma_{n-1}^{(F)}  + \sum_{m=0}^{n-3} \gamma_m^{(F)}  \gamma_{n-(m+3)}^{(F)}  \right],
\end{equation} 
with $\gamma_0^{(W)} = s^{(W)}$. This value of the initial coefficient is obtained directly from the difference of the initial coefficents $\gamma_0$ and $\gamma_0^{(F)}$. 
 
The above asymptotic condition on $h_k^{(W)}$ restricts the freedom in the choice of the integration constant in the solution \eqref{hkWbyparts} that we have obtained. Indeed, in the case without potential, the asymptotic equation \eqref{asymptotich} implies that $h_k^{(F)}$ behaves like $-ik$ up to subdominant terms for infinitely large values of $k$. Therefore, the exponential that accompanies $D_k^{(W)}(\eta_0)$ in Eq. \eqref{hkWbyparts} behaves as $e^{-2ik(\eta-\eta_0)}$ at this dominant order. The oscillatory behavior of this exponential is not compatible with an asymptotic series in inverse powers of $k$. Hence, in order to get an asymptotic behavior in consonance with  Eq. \eqref{asymptoticW}, we have to make the constant $D_k^{(W)}(\eta_0)$ equal to zero in this asymptotic regime. In cases in which there exist a (possibly unbounded) time $\eta_0$ at which $s^{(W)}/(2 h^{(F)})$ and all its derivatives with respect to $\tau_k$ vanish, it suffices to choose this time as the reference initial time and take the integration constant $C_k^{(W)}$ equal to zero. More generally, one can always set $D_k^{(W)}(\eta_0)$ to vanish (asymptotically), and fix our leading-order correction to $h_k$ so as to admit the following expansion when $k\rightarrow\infty$:
\begin{equation}\label{hkWfinal}
h_k^{(W)}(\eta)\sim - \sum_{n=0}^{\infty} \frac{d^n}{d \tau_k^n} \left(\frac{s^{(W)}(\eta)}{2h_k^{(F)}(\eta)} \right),
\end{equation}
This relation can be turned into an exact equality if the infinite sum of terms can be made meaningful for all $k$ by any means, as it would be the case e.g. if the sum converges.

By its very derivation, the asymptotic expansion \eqref{hkWfinal} that we have found must be consistent with Eq. \eqref{asymptoticW}. For illustrative purposes, let us indeed calculate the first terms in this asymptotic expression. On the one hand, the recurrence relation for $\gamma_n^{(W)}$ leads to the following lowest-order coefficients (all of them functions of the conformal time):
\begin{equation}\label{gammaWlow}
\gamma_0^{(W)} = s^{(W)}, \qquad \gamma_1^{(W)} = -\dot{s}^{(W)}, \qquad \gamma_2^{(W)} = \ddot{s}^{(W)} + 6s^{(F)}s^{(W)}.
\end{equation}
With this result and recalling that $\gamma_0^{(F)}=s^{(F)}$, we can compute the first terms in the asymptotic, inverse power series provided by Eq. \eqref{asymptoticW}:
\begin{equation}\label{hkWterms}
h_k^{(W)} \sim   -\frac{i}{2k} \left( s^{(W)} +\frac{i}{2k}\dot{s}^{(W)}- \frac{1}{4k^2} \ddot{s}^{(W)} - \frac{s^{(F)}s^{(W)}}{2k^2} \right) + \mathcal{O}(k^{-4}).
\end{equation} 
In this formula, we have not shown explicitly the dependence on the conformal time to simplify the notation. On the other hand, the expansion \eqref{asymptotich} for the case of a free evolution tells us that
\begin{equation}\label{asymphkFree}
\frac{1}{h_k^{(F)}} \sim \frac{i}{k} \left(1-\frac{s^{(F)}}{2k^2} \right) + \mathcal{O}(k^{-4}). 
\end{equation}
Using this asymptotic identity and the definition $d\tau_k=2h_k^{(F)}(\eta) d\eta$ in the expression \eqref{hkWfinal} of our solution $h_k^{(W)}$, one indeed recovers an expansion of the form \eqref{hkWterms}, confirming our fixation of this solution.

In summary, we have shown in this section  that, at least formally in the asymptotic region of large wavenumbers, it is possible to fix the leading-order correction caused by the potential in the function that determines the positive frequency solutions of the perturbations once this function is known for the free dynamics, and in this way proceed to select a vacuum state. 

\section{Conclusions}\label{conclusion}

Lately, there has been an increasing interest in cosmological models with scalar fields that present kinetically dominated phases in their evolution, both within classical General Relativity and in the context of quantum cosmology. The departures of this scenario from standard slow-roll inflation prevent one from employing the approximate analytical formulas for the power spectrum of the primordial perturbations that are valid in slow roll, complicating the calculations, that in most cases have to be done numerically. The complexity of these numerical calculations is much worst if the model takes into account the quantum behavior of the background, incorporating it by means of expectation values on the background geometry, as it happens in hybrid quantum cosmology. The background state evolves in time, requiring a calculation of the involved expectation values at each moment of the interval in which one studies the propagation of the perturbations. Moreover, the evolution of the quantum background state is typically not well controlled from a theoretical point of view in the presence of a scalar field potential, without a good knowledge of the properties of the eigenvalues of the associated quantum Hamiltonian that generates this background evolution [and even without a rigorous proof of its self-adjointness for generic potentials].  In hybrid LQC, for instance, this problem has been circumvented by considering background states that are highly peaked on effective trajectories, so that the discussed expectation values can be very well estimated by a direct evaluation on those trajectories. The price to pay is to renounce to consider more general quantum behaviors for the background. Furthermore, the fact that a quantum state is highly peaked with respect to certain variables does not immediately guarantee that the same happens for other non-linear functions of those variables, an issue that may cast some shades on the general validity of the evaluation on effective trajectories and that would be desirable to check by comparing this evaluation with a genuine quantum calculation.

In addition, the freedom to consider different potentials for the scalar field would make this panorama even more intricate. In this situation, an interesting possibility, that we have explored in this article, consists in approximating our description of the perturbations around the free case without potential, so that only the knowledge of the dynamics of this particular case is required in full detail. In order to consider the influence of the potential, it is necessary to include the corrections that its presence produces on this free dynamics. 

Here, we have analyzed these corrections at dominant, leading order. These corrections can be grouped in two sets, depending on whether they affect the propagation of the primordial perturbations in a strict sense or the natural vacuum state for those perturbations. The calculation of the primordial power spectrum actually requires the two pieces of information; the initial conditions provided by the vacuum and their evolution, that should eventually lead to the amplitude of the perturbations at the end of inflation (or at the end of the kinetically dominated phase, amplitude that would have to be evolved later on during the remaining interval until the end of inflation). In the models considered in this work, the corrections on the propagation of the perturbations are due to changes in the effective mass of the MS equations with respect to the free case. These changes are of three kinds. First, the explicit expression of the effective mass gets modified by the presence of a potential. Second, when this effective mass is given by quantum expectation values, the background state evolves with respect to the scalar field with a dynamics that gets contributions of the potential. And third, the change from the parameterization of this dynamics in terms of the homogeneous scalar field to the conformal time that appears in the MS equations is changed as well by the presence of the potential, in comparison with the situation with a free evolution. We have computed in detail these three types of changes and succinctly explained how the computation can be particularized to the cases of classical General Relativity, of hybrid LQC, and of hybrid quantum geometrodynamics. 

Concerning the changes in the vacuum of the perturbations, we have investigated a recent proposal for the choice of this state, which rests on a(n asymptotic) diagonalization of the contribution of the perturbations to the Hamiltonian constraint. This proposal leads to a selection of positive frequency solutions that are determined by a function $h_k$ which is a solution of certain Riccati equation and satisfies suitable asymptotic conditions. We have discussed the corrections to the function $h_k$ at leading order in the potential that are caused by the modification of the effective mass of the perturbations, mass that enters in the Riccati equation. In addition, we have studied the change in the asymptotic condition on $h_k$ owing to the modification of the effective mass. In this way, we have been able to give an expression for the leading-order 
contribution of the potential to the function $h_k$, which corrects the value of this function with respect to the case of a free scalar field. In this manner, we have derived the necessary tools to compute the leading-order effects of a small potential in the primordial spectrum of the perturbations during kinetically dominated regimes.

Let us notice that, in principle, the analysis that we have presented can be extended to cover higher-order corrections of the potential (although obviously at the cost of increasing computational difficulties). Our results open the possibility of studying issues such as the different effects that various potentials may produce in kinetically dominated epochs, if the leading-order approximation is good enough. In particular, this possibility may simplify the investigation of the consequences on the power spectra and the discussion about which potentials lead to an improved fitting between predictions and observations. Moreover, our results also facilitate the analysis of the quantum geometry effects on the primordial perturbations, which in models as those of effective LQC occur in kinetically dominated regimes. This could allow us to elucidate whether those quantum geometry effects leave an imprint on the spectra that can be observed in practice. Similarly, in principle it is now possible to discuss how much the vacuum state is changed by the presence of a potential during kinetically dominated epochs with important quantum geometry effects, such as those around the bounce in hybrid LQC. 

Another interesting question for future work is the application of our analysis to discriminate the effects of a fast-roll inflationary phase in models within General Relativity with respect to the quantum geometry effects that would be originated during a bounce, for instance in order to confirm some claims of recent studies in the literature about these two types of effects \cite{Bealast,Mercelast2}. Finally, the analysis that we have carried out here is sufficiently general so as to admit its extension to other quantum cosmology formalisms in which the net effect of the quantization is a change in the effective MS mass. As we mentioned in the Introduction, this is the case of the dressed-metric formalism for LQC \cite{AAN,AAN1,AAN2}. Another example is the hybrid quantization that one obtains in LQC when the background contribution to the Hamiltonian constraint is regularized in an alternative, recently proposed way \cite{Ma,DaporLieg,DaporLieg2,Wang1,Wang2,Agullo,Alex}. This regularization leads to an effective MS mass that is given again by a ratio of expectation values on a background state, but the operators that are involved are different from those considered in the standard representation used in hybrid LQC \cite{AlexhLQC,Alexmass}.

\section*{Acknowledgments}
This work was partially supported by Project. No. MICINN FIS2017-86497-C2-2-P from Spain (with extension Project. No. MICINN PID2020-118159GB-C41 under evaluation). B.E.N. acknowledges financial support from the Standard Program of JSPS Postdoctoral Fellowships for Research in Japan. The authors are grateful to L. Castell\'o Gomar, A. Garc\'{\i}a-Quismondo, L. Morala, and S. Prado for discussions.

\end{document}